\documentclass{article}
\usepackage[T1]{fontenc} % add special characters (e.g., umlaute)
\usepackage[utf8]{inputenc} % set utf-8 as default input encoding
\usepackage{ismir,amsmath,cite,url, amsfonts}

\usepackage{graphicx}
\usepackage{color}
\usepackage{booktabs}
\usepackage{pifont}% http://ctan.org/pkg/pifont
\usepackage[bookmarks=false,hidelinks]{hyperref}
\newcommand{\crmark}{\ding{55}}

\usepackage{lineno}
\title{Audio Conditioning for Music Generation \\via Discrete Bottleneck Features}

\multauthor
{Simon Rouard$^{1, 2}$ \hspace{0.5cm} Yossi Adi$^{1, 3}$ \hspace{0.5cm} Jade Copet$^1$ \hspace{0.5cm} Axel Roebel$^2$ \hspace{0.5cm} Alexandre Défossez$^4$} {
$^1$ FAIR Meta \hspace{0.3cm}
$^2$ IRCAM - Sorbonne Université \hspace{0.3cm}
$^3$ Hebrew University of Jerusalem \hspace{0.3cm}
$^4$ Kyutai \\
{\tt\small srouard@meta.com, alex@kyutai.org}
}

\def\authorname{S. Rouard, Y. Adi, J. Copet, A. Roebel, A. Défossez}

\begin{document}

\maketitle
\begin{abstract}
While most music generation models use textual or parametric conditioning (e.g. tempo, harmony, musical genre), we propose to condition a language model based music generation system with audio input. Our exploration involves two distinct strategies. The first strategy, termed textual inversion, leverages a pre-trained text-to-music model to map audio input to corresponding "pseudowords" in the textual embedding space. For the second model we train a music language model from scratch jointly with a text conditioner and a quantized audio feature extractor. At inference time, we can mix textual and audio conditioning and balance them thanks to a novel double classifier free guidance method. We conduct automatic and human studies that validates our approach. We will release the code and we provide music samples on \href{https://musicgenstyle.github.io}{\textcolor{blue}{musicgenstyle.github.io}} in order to show the quality of our model. 
\end{abstract}

\section{Introduction}

% Generative models recently became really good in various domains such as text, image, video, speech or audio. 
In the field of music generation, prior research has predominantly focused on producing brief musical segments \cite{gansynth, crash}, MIDI generation \cite{midinet}, while generating long and coherent waveforms (around 30 seconds) has only recently been tackled \cite{musiclm, musicgen, stableaudio}. Specifically, most of these recent models have been designed to perform text-to-music generation, providing a fascinating tool for creators. Other types of high-level conditioning have been used in previous work such as tempo, harmony \cite{mustango}. For lower-level and aligned conditioning, the authors of \cite{musicgen} use melody, while \cite{cocomulla} uses chords, piano rolls, or the drum stem. However, music is hard to describe textually and the scarcity of text-music pair datasets makes it challenging to generate music in the style of a specific artist or song, since the artist is probably not represented in the training dataset. Then a common use case would be to generate music in the style of a reference segment. This gives more control to the user since they do not have to find a textual prompt that describes the music they want to generate. 

In the computer vision domain, the authors of \cite{image_textual_inversion} introduced textual inversion to extract visual concepts that can then be used to generate new images with a text-to-image model. Given a few images (3-5) of a concept or object, one sets them as outputs of a frozen text-to-image model with a randomly initialized learnable text embedding. Backpropagating the generative model loss on the text allows to learn new "pseudowords" in the textual embedding space of the model that match the common concept depicted on the images. One can then compose this learnt pseudoword $S^*$ in a textual prompt to generate an image of the learnt concept (for instance "a painting of $S^*$ in the style of Picasso"). 

We first adapted this method by using the text-to-music model MusicGen \cite{musicgen}, using crops of a song to depict a concept, and optimizing the cross-entropy loss of the music language model. 
This approach does not need to retrain a model from scratch. However, its inference is very slow since it requires hundreds of optimization steps of the textual prompt, including gradient computation through the language model, before generating music. 

To tackle this issue, we present another method where we design a style conditioner module that we jointly train with a text-to-music MusicGen model \cite{musicgen}. This style conditioner takes a few seconds of audio and extracts features out of it. As a result this new model can generate music using two modalities as input: waveforms and textual descriptions. Our conditioning is high level even if it can retain some lower level content such as melodic patterns or rhythm. 
Designing this style conditioner is challenging as we need to extract enough features to have a meaningful conditioning but not too much, to prevent the generative model to copy and loop the conditioning audio. We thus need to introduce and tune information bottlenecks in our conditioning module.
%, e.g. using Residual Vector Quantization \cite{soundstream} and downsampling.
Our contributions are the following:

1) We adapt the textual inversion method of \cite{image_textual_inversion} to a pretrained text-to-music MusicGen model. This allows to perform audio conditioning for music generation without training a model from scratch. 

2) We present our style conditioner method which is based on a frozen audio feature extractor (Encodec \cite{encodec}, MERT \cite{mert} or MusicFM \cite{musicfm}) followed by a transformer encoder \cite{transformer}, Residual Vector Quantizer (RVQ) \cite{soundstream} and temporal downsampling. The number of residual streams used by RVQ is adjustable at inference time which gives the user the ability to change the strength of the style conditioning. To our knowledge, we are the first to explore this approach for music generation. 
%this is the first work that explores this. 

3) Since the model is trained with both textual and audio conditioning inputs, we can combine both to generate music. However, audio contains much more information, so that text is ignored by the model at inference. We propose to balance them with a new double classifier free guidance \cite{cfg} which is a general method for merging conditions with various degrees of information. 

4) We introduce novel objective metrics for style conditioning, based on nearest neighbors search in the latent space, validated with human evaluations.

We compare our method to baselines which are: a MusicGen trained with CLAP embeddings \cite{clap} as conditioning, a text-to-music MusicGen used with text prompts, and a MusicGen model without conditioning used in continuation mode. We perform as well some ablation studies in order to justify the architecture of our style encoder. Based on results, we show the practicality of our methods and the musical quality of the generated music. 

\section{Related work}
\subsection{Generative models for music}
Music generation models can be categorized into two types: autoregressive models and non autoregressive ones. 

\noindent\textbf{Autoregressive} ones are motivated by the successful work done in natural language modeling. Recent successful models use a compression model taking the form of a multi stream quantized autoencoder \cite{soundstream, encodec} in order to convert audio into $K$ parallel discrete streams of tokens. The $K$ streams are obtained by performing Residual Vector Quantization (RVQ) \cite{soundstream} on the latent space of an autoencoder, making the first stream contain coarse information and following ones refine the approximation of the latent space. 
Then, an autoregressive transformer \cite{transformer} is used to model these audio tokens.  MusicLM \cite{musiclm}  and MusicGen \cite{musicgen} are built on this principle. MusicLM uses a multi-stage approach with different models to predict the $K$ streams, while MusicGen models them in parallel using a delay pattern \cite{kharitonov-etal-2022-text, musicgen}. 

\noindent\textbf{Non-autoregressive} models such as AudioLDM2 \cite{audioldm2}, MusicLDM \cite{musicldm}, and Stable Audio \cite{stableaudio}, are latent diffusion models operating in the latent space of a continuous variational autoencoder. Some other models use cascaded diffusion such as Noise2Music \cite{noise2music} to progressively increase the sampling rate of the audio. Mo\^usai \cite{mousai} uses a first diffusion model to compress the music and a second one to generate music from this representation and textual descriptions. MusTango \cite{mustango} uses a latent diffusion model conditioned on textual description, chord, beat, tempo and key. 
Jen-1 \cite{jen1} combines a diffusion model and a masked autoencoder trained with multi-tasks objectives. It can perform music generation, continuation and inpainting. A second version \cite{jen1composer} uses source separation\cite{demucs} over their dataset to allow the user to generate and edit music stem by stem. 
VampNet \cite{vampnet} is a masked modeling approach to music synthesis that uses masking at training and inference time in order to generate discrete audio tokens.
MAGNeT \cite{magnet} is based on the same masking principle. It can also combine autoregressive and masking to reach the same quality as the autoregressive baseline (MusicGen) but with a 7x faster inference. In MeLoDy \cite{melody}, a language model is used to model coarse semantic tokens and a dual path diffusion model is then used for acoustic modeling. The authors claim faster than real time generation. 

\subsection{Jointly trained conditioners for music generative models}
Regarding the conditioning, most of the models focused on text-to-music \cite{musiclm, musicldm, musicgen, noise2music, mousai, jen1}. Since pairs of text-music data are rare, most models use a pre-trained contrastive text-music model such as CLAP \cite{clap} or MuLan \cite{mulan}, to condition their text-to-music models. Then, massive amount of non-annotated audio data can be used at training time and text is used at inference time. However, these text-to-music models never exploit the fact that audio can be used as conditioning. For other types of conditioning, MusTango \cite{mustango} is trained with text, beat tempo, key and chords as conditioning, StableAudio \cite{stableaudio} takes timing embeddings to control the length and structure of the generated music. Some models generate stems while being conditioned on other stems. For instance, SingSong \cite{singsong} generates musical accompaniments from singing and Jen-1 Composer \cite{jen1composer} handles multi-track music generation on 4 different stems (bass, drums, instrument and melody). MusicGen \cite{musicgen} and Music ControlNet \cite{musiccontrolnet} can handle melody as conditioning and the latter can also use dynamics and rhythm. Both papers use chromagrams extraction for melody conditioning. %More widely, in the field of audio generation, the authors of \cite{image2audio} perform image-to-audio generation. %\textcolor{red}{maybe should speak about high-level/low-level conditioning? }

\subsection{Conditioning a pretrained generative model}

\noindent\textbf{With finetuning}:
In Coco-Mulla \cite{cocomulla}, the authors use parameter-efficient fine-tuning (PEFT) to specialize a text-to-music MusicGen model on chords and rhythm. They finetune on a number of parameter that is 4\% the amount of parameters of the original network with only 300 songs.
Music ControlNet \cite{musiccontrolnet} is a finetuned text-to-music diffusion model that operates in the spectral domain. The finetuning strategy comes from the text-to-image method ControlNet \cite{controlnet} and allows to handle melody, dynamics and rhythm conditioning. The pixel-level control that allows ControlNet on images gives a pixel-level control on the mel-spectrogram. 

\noindent\textbf{Without finetuning}:
In \cite{music_textual_inversion}, the authors use AudioLDM \cite{audioldm2} as a backbone to perform textual inversion \cite{image_textual_inversion}. For each textual inversion they use a group of 5 excerpts of 10 seconds. They also try an experiment where they optimize the pseudoword $S^*$ as well as the diffusion neural network which gives them better results.
In \cite{controllable_apple}, the authors use a diffusion model trained on musical data with no conditioning and perform various interactive tasks at inference which are infilling, continuation, transition (smooth a transition between two songs) and guidance. The one that is the most similar to our audio conditioning is the guidance where the diffusion model is guided by the PaSST classifier \cite{passt} embedding of an audio prompt. However the model only generates 5 seconds excerpts of music. 
Some other papers involve new control with no finetuning such as in \cite{zeroshot} or DITTO \cite{ditto} where the authors use a pre-trained text-to-music diffusion model and control its inference by optimizing the initial noise latent. In SMITIN \cite{smitin}, the authors control a pretrained MusicGen model by steering the attention heads in the direction that maximizes the probability of some features. %They apply this to several applications that are inpainting, outpainting, looping, intensity, melody and musical structure control. 

\section{Textual Inversion Method}
\begin{figure}[t]
    \centering
    \scalebox{0.7}{
    \includegraphics[width=0.5\textwidth]{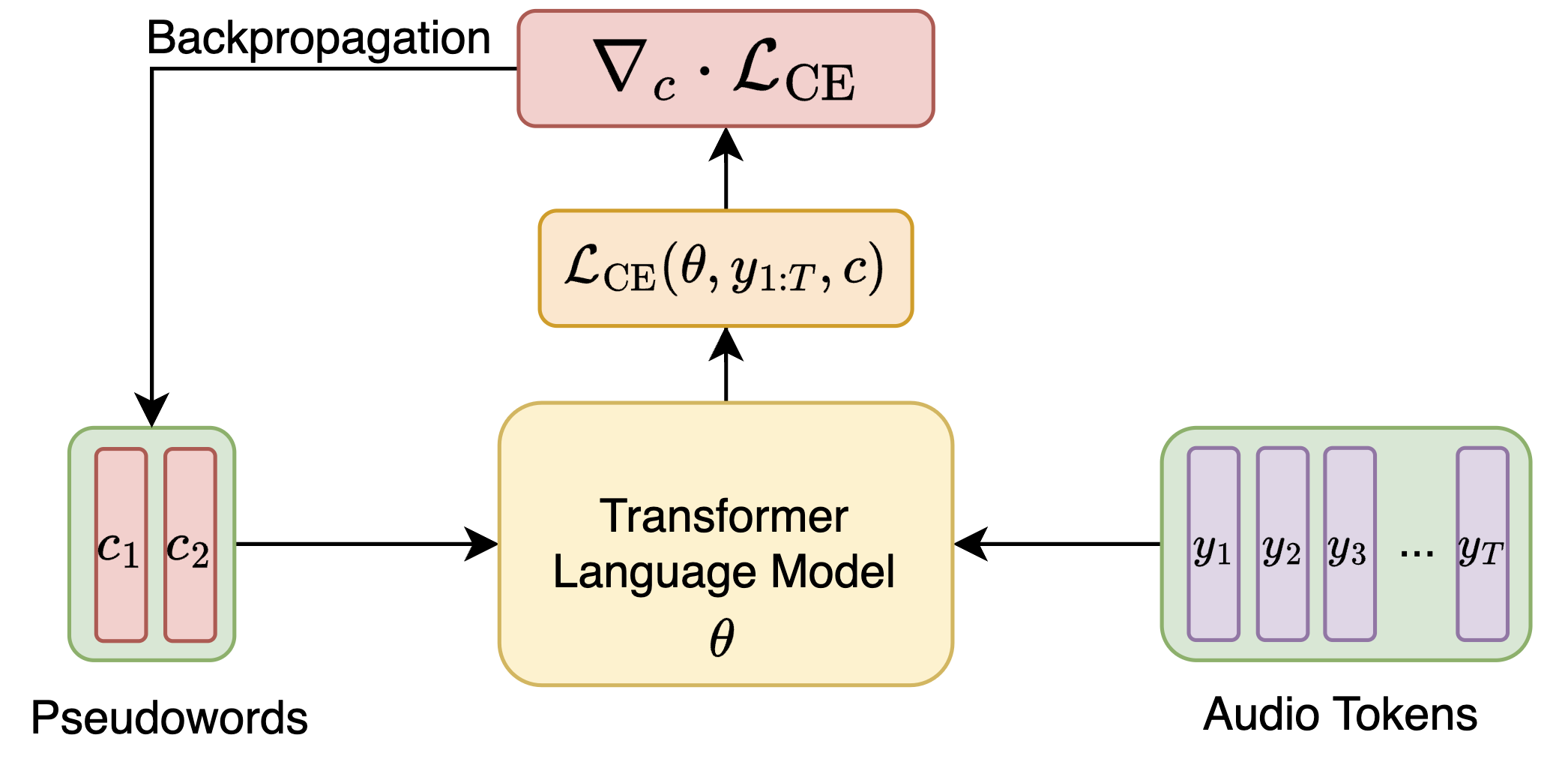}}
    \caption{An overview of the Texual Inversion method based on the pretrained text-to-music MusicGen}
    \label{fig:textual_inv}
\end{figure}

We first present our textual inversion method in the case of autoregressive modeling (see Fig.~\ref{fig:textual_inv}). It is based on previous work in the image domain \cite{image_textual_inversion} with diffusion models. 

Autoregressive modeling aims to estimate the conditional distribution of the next token $y_t$ given the preceding tokens $y_{<t}$ and a conditioning context $c$, such as a textual embedding. In the framework of transformer decoder neural networks parameterized by $\theta$, denoted as $p_\theta$, this conditional distribution is typically modeled as a product of individual probabilities:
\begin{equation}
    p_{\theta}(y_{1:T} | c) = \prod_{t=1}^{T} p_{\theta}(y_t | y_{<t}, c)
\end{equation}
Here, $y_{1:T}$ represents the sequence of tokens, and $p_{\theta}(y_t | y_{<t}, c)$ denotes the probability of observing token $y_t$ given the preceding tokens and the conditioning context. During training, with a given sequence $y_{1:T}$ and its associated textual description $c$, we compute the cross-entropy loss:
\begin{equation}
    \mathcal{L}_\mathrm{CE}(\theta, y_{1:T}, c) = - \sum_{t=1}^{T} \log p_{\theta}(y_t | y_{<t}, c)
\end{equation}
It is minimized by taking a gradient descent step on $\nabla_{\theta} \mathcal{L}_\mathrm{CE}(\theta, y_{1:T}, c)$.
This loss quantifies the dissimilarity between the predicted conditional distribution and the true distribution of the next token, serving as the optimization objective for training autoregressive models. 

For the textual inversion method, we take a pretrained text-to-music MusicGen for the transformer decoder. We initialize the textual embedding (for instance with the textual embedding of the word "music") $c$. Given a song $Y$, we cut it into random chunks $\{y_{1:T}^{i}\}_i$ and optimize the textual embedding $c$ by taking successive gradient steps on $\nabla_{c} \mathcal{L}_\mathrm{CE}(\theta, y_{1:T}^{i}, c)$. After a few hundreds iterations the learnt $c$ is fed into the text-to-music MusicGen model to generate a song in the style of $Y$.

\section{Style Conditioning Method}
\begin{figure*}[h]
    \centering
    \scalebox{0.6}{
    \includegraphics[width=1.0\textwidth]{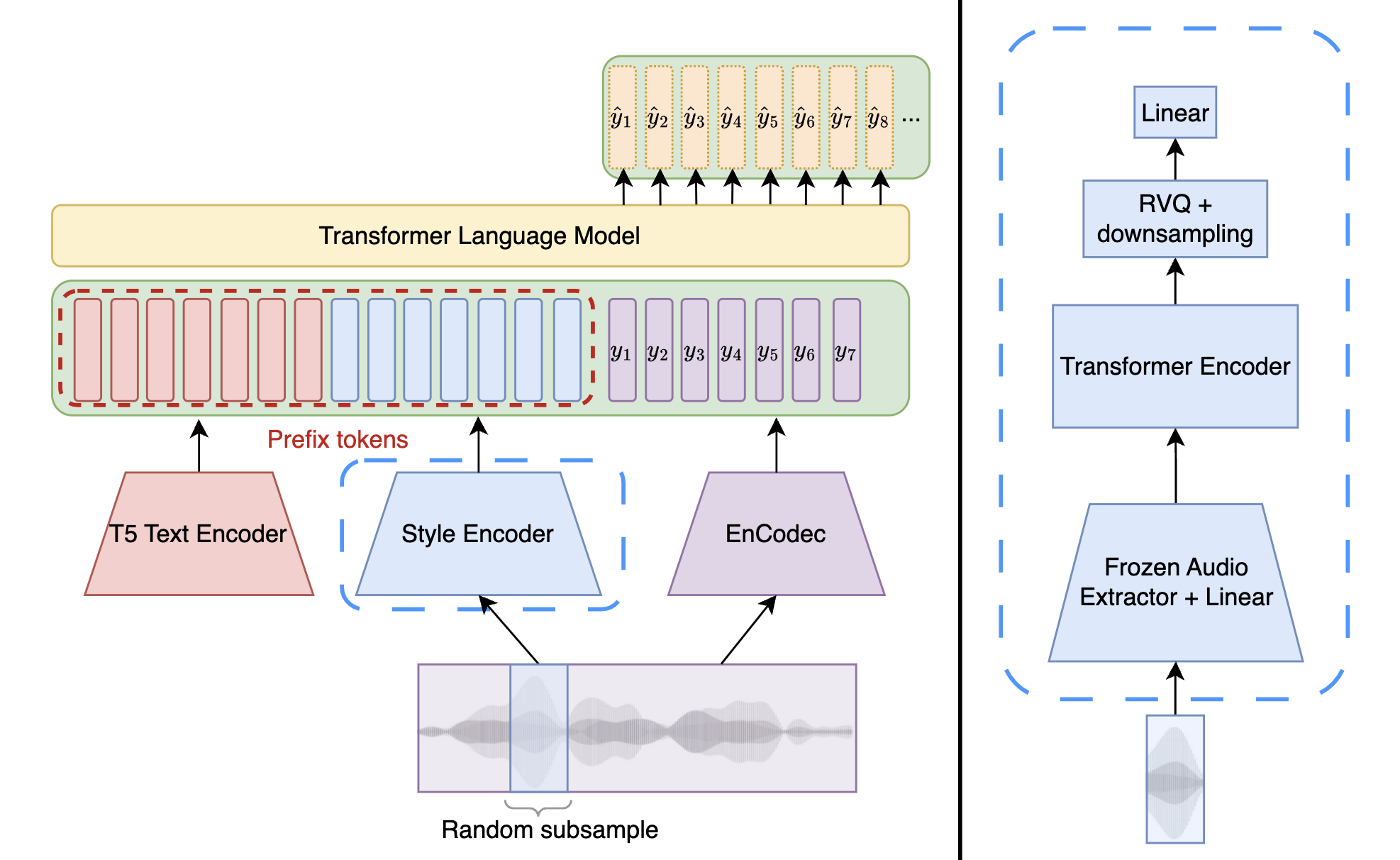}}
    \caption{An overview of the general architecture. Text conditioning and style conditioning are provided to the model as a prefix. On the right we present the style conditioner.} 
    \label{fig:musicgen_style}
\end{figure*}

\subsection{General Architecture}
The general architecture, depicted on the left of Fig.~\ref{fig:musicgen_style}, is based on the text-to-music model MusicGen \cite{musicgen} with the addition of a style conditioner that is jointly trained with the language model. At train time, a 30 seconds music excerpt paired with a textual description is input to the model. The textual description is fed into a frozen T5 tokenizer and transformer encoder \cite{t5}. The style encoder takes a random subsample (between 1.5 and 4.5 seconds) of the input audio and encodes it. 
The text and style latent representations are both projected with linear layers to have the same dimension as the transformer language model, and provided as prefix to the sequence to model.

The input audio is encoded by a pretrained EnCodec \cite{encodec} model and the language model is trained in a autoregressive manner with a cross-entropy loss. In addition, the tokens that correspond to the random subsample fed into the style encoder are masked in the loss, as we noticed this reduces the tendency of the model to just copy the style audio input.
At inference time, we can use text or/and a short excerpt of music as a conditioning to generate music. 

\subsection{Architecture of the Style Conditioner}

Our style conditioner is designed with bottlenecks (RVQ \cite{soundstream} and downsampling) to prevent transmitting all the information of the conditioning audio excerpt to the model. Without these bottlenecks, the generative models retrieves easily the excerpt and copies it (see the ablation study in Sec.~\ref{sec:ablation}). The style conditioner depicted on the right of Fig.~\ref{fig:musicgen_style} takes an audio input of length 1.5 to 4.5 seconds, passes it through a frozen feature extractor followed by a trainable transformer encoder and a residual vector quantization (RVQ) module with 6 codebooks. After quantization, we downsample on the temporal axis to obtain a conditioning with a 5Hz frame rate which gives a similar length as a text description (8 to 25 tokens). 
Finally a linear layer outputs the same dimension as the language model. 

The candidates for the audio encoder are a Encodec followed by trainable embeddings for each codebook that are summed, a transformer based music foundation model from \cite{musicfm} (we now name it MusicFM for the rest of the paper) where the authors claim state of the art on several downstream tasks specific to music information retrieval and a MERT model \cite{mert}, a transformer based music model trained in a self-supervised manner. The first one has a frame rate of 50Hz and 60M parameters, the second one has a frame rate of 25Hz and 620M parameters and the third one has a frame rate of 75Hz and 95M parameters

At training time, we use dropout on the conditioning, keeping both conditions 25\% of time, one of the two conditions 25\% of time for each (no text or no style) or no condition 25\% of time. There is also a dropout on the number of the codebooks used by the RVQ of the style conditioner: at each step of the training, the number of used codebooks is uniformly sampled between 1 and 6. Then, at inference time, we can control the bottleneck of the style conditioner. Setting the number of codebooks to 1 gives more flexibility to the generative model whereas using 6 levels of quantization constraints it more. In practice, this means that music generated with 6 streams of quantization will sound more similar to the input condition compared to music generated with 1 stream of quantization.

\begin{table*}[t]
    \centering
    \resizebox{\textwidth}{!}{
    \begin{tabular}{l |ccccc|ccc}
    \toprule
         Model & FAD$_{vgg} \downarrow$ & KL $\downarrow$ & CLAP $\uparrow$ & $\text{KNN}_\text{common}$ $\uparrow$ & $\text{KNN}_\text{overfit}$ $\downarrow$ & OVL $\uparrow$ & SIM $\uparrow$ & VAR $\downarrow$ \\ \hline
    Textual Inversion  & 6.07 & 0.55 & 0.20 & 0.20 &  0.14 & 78.11 $\pm$ 0.93 & 61.78 $\pm$ 1.06 & 69.53 $\pm$ 1.44\\
    MusicGen Continuation  & 1.22 & 0.51 & 0.30 & 0.26 & 0.17 & 83.95 $\pm$ 0.83 & 73.38 $\pm$ 0.97 & 77.24 $\pm$ 1.29\\
    MusicGen w. audio CLAP & 0.96 & \textbf{0.43} & \textbf{0.31} & 0.09 & 0.02 & 84.76 $\pm$ 0.93 & 62.37 $\pm$ 1.04 & 68.58 $\pm$ 1.42 \\ \hline
    Our Model w. EnCodec, 2 RVQ & \textbf{0.85} & 0.49 & 0.29 & 0.23 & 0.12 & 83.41 $\pm$ 1.04 & 72.16 $\pm$ 0.93 & 72.39 $\pm$ 1.33 \\
     \bottomrule
    \end{tabular}}
    \caption{Comparison with baselines. The $\text{KNN}_*$ metrics, introduced in Sec. \ref{sec:objective_metrics}, measure how close the generation is from the style condition, yet different from the matching ground truth. Those are completed with the subjective evaluations from Sec. \ref{sec:subjective_metrics}. While using MusicGen for continuation matches well to the style audio, it has limited variation. Using a CLAP audio encoder as conditioning does the opposite, while using our style encoder gets the right balance between the two.}
    \label{tab1}
\end{table*}

\subsection{Double Classifier Free Guidance}

When doing next token prediction, let’s denote $l_{\text{style, text}}$ the logits of the model conditioned on style and textual description. 
Classifier free guidance  \cite{cfg} consists of pushing the logits in the direction predicted with the conditioning, to increase its importance:
\begin{equation}
    %l_{\text{CFG}}(\emptyset \Rightarrow (\text{style, text})) = l_{\emptyset} + \alpha (l_{\text{style, text}} - l_{\emptyset})
    l_{\text{CFG}} = l_{\emptyset} + \alpha (l_{\text{style, text}} - l_{\emptyset}), \text{with } \alpha>1,
\end{equation}

typically, $\alpha=3$ is used in previous work \cite{musicgen}. 

When generating music with a textual description that contradicts the audio of the style conditioning, we observe that the description is ignored by the model. This is explained by the fact that audio is more informative conditioning compared with the text, so that the model weights it more during training.
To counteract this effect, we introduce a \emph{double classifier free guidance} in which we iterate the CFG formula: we first push from style only to style and text and we then push these logits a second time from no conditioning. 
\begin{equation}
\label{eq:double_cfg}
    l_{\text{double CFG}} = l_{\emptyset} + \alpha [l_{style} + \beta(l_{text, style} - l_{style}) - l_{\emptyset}]
\end{equation}

We retrieve the simple CFG with $\beta = 1$. For $\beta > 1$, we boost the importance of the text conditioning (see Sec.~\ref{sec:cfg}).

\subsection{Objective Metrics}
\label{sec:objective_metrics}
The difficulty with generating samples in the same style of a song is that we want to generate something that is similar enough but not too close. This is something that can be subjectively evaluated. For easing the comparison of various approaches and hyper parameters, we also introduce a novel set of objective metrics.

\noindent\textbf{Nearest Neighbours in Common}: Let’s note $x_C \in \mathbb{R}^{D \times T}$ ($D=1$ for mono music) the audio that we input in the style conditioner and $x_G \in \mathbb{R}^{D \times T’}$ the generated sequence. We use an encoder $E: \mathbb{R}^{D \times T} \rightarrow \mathbb{R}^{N}$ which outputs a single vector whatever the input length $T$ is. In practice, this is done by taking a MusicFM model and averaging on the time dimension. 
Then, for each song of our valid and test sets, we cut it into chunks of 30 seconds and store the embeddings $\{E_{i,j}\}$, $i$ being the index of the song and $j$ the chunk number. For $E_C = E(x_C)$, we compute the cosine similarities $\cos(E_C, E_{i, j}), \forall i, j$ and the set of its $K$ nearest neighbors: $\{i_1^C, ... i_K^C\}$. We do the same for $E_G = E(x_G)$ and obtain a set of $K$ values $\{i_1^G, ... i_K^G\}$. We then have found the nearest songs in the dataset. We define our metric $\text{KNN}_{\text{common}}(x_C, x_G)$ for a song $x_G$ that has been generated after being conditioned by $x_C$:
\begin{equation}
    \text{KNN}_{\text{common}}(x_C, x_G) = \frac{|\{i_1^C, ... i_K^C\} \cap \{i_1^G, ... i_K^G\}|}{K} \in [0, 1].
\end{equation}
The intuition behind this metric is that a model performs well at recreating a song in the style of another if the generated song and its conditioning audio have embeddings that are close enough to share neighbors in the dataset. However, if a model copies the conditioning (i.e. $x_G \approx x_C$) the metric will tend to $1$, we thus need a second metric to avoid $x_G$ and $x_C$ being too similar.

\noindent\textbf{G is the Nearest Neighbor of C}: We want $E_G$ and $E_C$ to be close while being different. One way to be sure that the corresponding audios are not too similar is to check that if we add $E_G$ to the set of embeddings $\{E_{i,j}\}$, $E_G$ is not the nearest neighbor of $E_C$. Assuring that another song from the dataset is closer to the conditioning means that the model is creative enough and not just copying its input. 
Formally, denoting $\{E_{\cup}\} = \{E_{i,j}\} \cup \{E_G\}$, we define

%\resizebox{0.95\columnwidth}{!}{
\begin{equation}
\mathrm{KNN}_{\mathrm{overfit}}(x_C, x_G) =  
    \begin{cases}
    \displaystyle
1 \, \text{if }\, \underset{E \in \{E_{\cup}\}}{\mathrm{argmax}}\left[\cos(E_C, E)\right] {=} E_G \\
0 \,\, \text{otherwise.}
\end{cases}
\end{equation}
%}
For our evaluations, we take 1000 samples of 3 seconds $x_C$ from our test set, generate the corresponding $x_G$ and average the two KNN metrics. Intuitively, the two metrics are positively correlated, but for a similar value for $\text{KNN}_{\text{common}}$ we will favor the model that minimizes $\text{KNN}_{\text{overfit}}$.

\noindent\textbf{Other Objective Metrics} To evaluate the quality of the generated music, we also use the official implementation of the Fr{\'{e}}chet Audio Distance defined in \cite{fad} that uses a VGGish model, the KL-divergence based metric introduced in \cite{musicgen} that computes the KL-divergence on the probabilities of the labels of a pretrained audio classifier between the conditioning and the generated music. 
We noticed that a high FAD (> 2) often indicates a poor quality of the generated samples. The CLAP score \cite{clap, musicgen} computes the cosine similarity between the description and the audio embeddings obtained with the CLAP model. A higher score indicates that the generated audio aligns well with the textual description of the conditioning audio. 

\subsection{Human studies metrics}
\label{sec:subjective_metrics}
We follow a similar protocol as in \cite{musicgen} for the human studies. We ask human raters to evaluate three different aspects of the generated audio:
(1) How would you rate the overall quality of this excerpt [OVL]?
(2) Without considering audio quality, how similar are these two excerpts in terms of style [SIM]?
(3) Without considering audio quality, how likely do you think these two excerpts are from the same song [VAR]?

We believe that the SIM and VAR scores are the subjective versions of $\text{KNN}_{\text{common}}$ and $\text{KNN}_{\text{overfit}}$. 
\section{Experimental results}

\begin{table*}[ht]
    \centering
    \resizebox{0.85\linewidth}{!}{
    \begin{tabular}{cc |ccccc|ccc}
    \toprule
         Feat. Ext. & Quant. & FAD$_{vgg} \downarrow$ & KL $\downarrow$ & CLAP $\uparrow$ & $\text{KNN}_\text{common}$ $\uparrow$ & $\text{KNN}_\text{overfit}$ $\downarrow$ & OVL $\uparrow$ & SIM $\uparrow$ & VAR $\downarrow$ \\ \hline

    MERT & 1 & 0.78 & 0.50 & 0.29 & 0.19 & \textbf{0.06} & 84.07 $\pm$ 0.93 & 70.27 $\pm$ 1.22 & \textbf{69.69 $\pm$ 1.31}\\
    MERT & 2 & 0.75 & 0.47 & 0.30 & 0.24 & 0.13 & 84.14 $\pm$ 0.96 & 72.53 $\pm$ 1.05 & 72.81 $\pm$ 1.21\\
    MERT & 4 & 0.75 & \textbf{0.45} & \textbf{0.31} & \textbf{0.29} & 0.18 & \textbf{84.32 $\pm$ 1.04} & \textbf{74.15 $\pm$ 0.96} & 75.12 $\pm$ 1.35\\ \hline

    EnCodec & 2 & 0.85 & 0.49 & 0.29 & 0.23 & 0.12 & 84.02 $\pm$ 0.89 & 72.69 $\pm$ 0.91 & 72.47 $\pm$ 1.28\\ \hline

    MusicFM & 2 & \textbf{0.70} & \textbf{0.45} & \textbf{0.31} & 0.28 & 0.16 & 84.45 $\pm$ 1.09 & 73.01 $\pm$ 0.95 & 74.01 $\pm$ 1.36\\
     \bottomrule
    \end{tabular}}
    \caption{Comparison between the 3 feature extractors. The human studies correlate well with the KNN metrics.
    As expected, using coarser quantization of the style features leads to more variations in the generated audio.
    Self-supervised encoder like MERT and MusicFM outperforms low level acoustic models like EnCodec.
    }
    \label{tab2}
\end{table*}

\begin{table}
    \centering
    \resizebox{\linewidth}{!}{
    \begin{tabular}{l |ccccc}
    \toprule
         Model & FAD$_{vgg} \downarrow$ & KL $\downarrow$ & CLAP $\uparrow$ & $\text{KNN}_\text{common}$ $\uparrow$ & $\text{KNN}_\text{overfit}$ $\downarrow$ \\ \hline
    
    Our Model & \textbf{0.75} & \textbf{0.45} & \textbf{0.31} & 0.29 & 0.18  \\
    Smaller Transformer & 0.98 & 0.48 & 0.29 & 0.24 & 0.13 \\
    No Transformer & 2.92 & 0.96 & 0.13 & 0.01 & 0.0 \\
    No Masking of the loss & 1.11 & 0.53 & 0.29 & 0.22 & 0.30 \\
    %No RVQ  & 0.81 & \textbf{0.44} & 0.31 & 0.30 & 0.19 \\
     \bottomrule
    \end{tabular}}
    \caption{Ablation Study on our model with a MERT feature extractor with 4 quantization streams.}
    \label{tab:ablation}
\end{table}

\subsection{Hyperparameters for the textual inversion}
For the textual inversion method we test different parameters sets and retain these ones: we use a 12 tokens sentence for initialization, a batch size of 8 with 5 seconds segments randomly sampled from a 30 second excerpt with 200 optimization steps, a learning rate of 0.025 with a vanilla Adam optimizer. Finally the main issue that we encounter with this method is its instability. It is hard to find a set of hyperparameters that works well for any song. Some songs seem to be easier to invert for different sets of hyperparameters. For some song, we never achieve to obtain hearable music as the result suffers from glitches, and tempo instabilities.  
Finally, it seems beneficial to augment the length of the text embedding, as well as performing the inversion over chunks taken from a 30 seconds excerpt rather than the entire song. The results are shown in Tab.~\ref{tab1}. 

\subsection{Hyperparameters for the style conditioner}
All the models that we train are medium size (1.5B parameters) MusicGen models built on top of the 4 stream 32kHz music version of EnCodec \cite{encodec}. All models are trained for 400K steps on 64 V100 GPUs with the AdamW optimizer using $\beta_1=0.9$, $\beta_2=0.95$, a batch size of $192$, and music sequences of $30$ seconds. For the style conditioner, excerpts between 1.5 and 4.5 seconds are subsampled from the original sequence, the transformer encoder has 8 layers, 8 heads, a dimension of 512 and is non-causal, the residual vector quantizer has a codebook size of 1024, 6 streams and a variable number of streams is sampled at each training step, hence allowing the language model to train on all the levels of quantization. The style tokens are downsampled to 5Hz. All our evaluations are done on 1000 samples of the test set. Similarly to the MusicGen Melody model, both the textual description and the style condition are provided as prefix to the language model. 

\subsection{Datasets}
We use 20K hours of licensed music as in \cite{musicgen}. The training dataset is composed of 25K and 365K songs from the ShutterStock and Pond5 music data collections, as well as 10k tracks of an internal dataset. 
Each song comes with textual description, and is downsampled to 32kHz mono.

\subsection{Comparison with baselines and model selection}
Apart from the closed-source model udio \cite{udio}, there is no other audio conditioned music generative model. We use as a baseline a MusicGen model in the continuation setting: given 3 seconds of music, we ask MusicGen to continue the music with no textual prompt. For the second one we train a MusicGen model with a pretrained CLAP audio encoder \cite{clap} as conditioning, also taking 3 seconds of audio as input. In Tab.~\ref{tab1}, we compare these two baselines with our model with the EnCodec feature extractor for the style conditioner, a quantization level of 2 and with a textual inversion model. We notice that the FAD correlates well with the quality metric (OVL) since the textual inversion model has the worst OVL and FAD scores. Thus excluding this approach, we observe that the $\text{KNN}_\text{common}$ and the SIM metrics ranks the models in the same orders as well as the $\text{KNN}_\text{overfit}$ and VAR metrics. 

Regarding the baselines, the textual inversion method provides results of poor quality (FAD). The continuation method provides music that has a high similarity to the conditioning (high $\text{KNN}_\text{common}$ and SIM) but that is too similar to it (high $\text{KNN}_\text{overfit}$ and VAR). However, the CLAP conditioning captures a more vague style of the conditioning and generates music that is too far from it (low $\text{KNN}_\text{common}$, $\text{KNN}_\text{overfit}$, SIM and VAR). Our model with the EnCodec feature extractor and 2 levels of quantization strikes the right balance between these two baselines.
 
In order to strengthen our claim that our KNN metrics correlates well with human perception of closeness between musical excerpts, we showcase a second study in Tab.~\ref{tab2}. In this study we compare the metrics of the MERT feature extractor with 3 quantization levels 1, 2, 4 (we recall that the models can go up to 6) as well as the EnCodec and MusicFM feature extractors with a quantization level of 2. All models generate music of similar quality (FAD and OVL). We notice that when the bottleneck is larger (i.e. when the quantization level is higher), the $\text{KNN}_\text{common}$ augments. This follows the intuition that if the conditioner transmits more information to the language model, the generated music will be closer to the input condition. The models follows similar orders for $\text{KNN}_\text{common}$ and SIM as well as for $\text{KNN}_\text{overfit}$ and VAR.

\subsection{Ablation Study}
\label{sec:ablation}
We perform an ablation study in Tab.~\ref{tab:ablation} on the components of the style conditioner with MERT as a feature extractor, and 4 RVQ streams. When reducing the size of the transformer encoder from 8 layers and 512 dimensions to 4 layers and 256 dimensions, the quality of the generated audio is worse. When removing the transformer encoder, the model generates audio that is far from music (high FAD). When we don't mask the music that is input to the style conditioner in the cross-entropy loss at training time, the audio quality is slightly worse and the model generates music that is too close to the conditioning and tends to loop. The very high $\text{KNN}_\text{overfit}$ indicates it since for a $\text{KNN}_\text{common}$ lower than the best model the $\text{KNN}_\text{overfit}$ is twice its value. 

\subsection{Tuning the Classifier Free Guidance}
\label{sec:cfg}
\begin{table}
    \centering
    \resizebox{0.35\textwidth}{!}{
    \begin{tabular}{ccc|ccc}
    \toprule
         Type & $\alpha$ & $\beta$ & FAD$_{vgg} \downarrow$  & CLAP $\uparrow$ & $\text{KNN}_\text{common}$ $\uparrow$ \\ \hline
    
    % AudioLDM2 & 347M & 3.27 & 1.19 & 34 & 84.06 $\pm$ 0.85 & \\
    No CFG & \crmark & \crmark & 1.54 & 0.25 & 0.088  \\
    simple & 3 & \crmark & 0.92 & 0.28 & 0.162  \\
    double & 3 & 3 & 0.80 & 0.35 & 0.123  \\
    double & 3 & 4 & 0.78 & 0.37 & 0.104  \\
    double & 3 & 5 & 0.84 & 0.37 & 0.095  \\
    double & 3 & 6 & 0.97 & 0.38 & 0.081  \\
    %Cross-Attention for description & 1.31 & 0.56 & 0.28 & 0.196 & 0.175 \\
     \bottomrule
    \end{tabular}}
    \caption{
    Classifier Free Guidance parameters tuning.
    Larger $\beta$ from \eqref{eq:double_cfg} leads to increasing the importance of the text conditioning (given by the CLAP score), and decreasing the similarity to the style audio, given by $\text{KNN}_\text{common}$.}
    \label{tab:cfg}
\end{table}
When style and text conditioning are both used and are not consistent, it is necessary to use double CFG instead of simple CFG so that the text is not ignored. To tune the parameters $\alpha, \beta$ of the double classifier free guidance given by \eqref{eq:double_cfg}, we rely on the following protocol. For 1000 samples of our test set, we randomly shuffle text descriptions and generate music while conditioning both on text and music. We track the FAD \cite{fad}, the $\text{KNN}_\text{common}$ and the CLAP score. In Tab~\ref{tab:cfg} we observe the intuitive fact that the $\text{KNN}_\text{common}$ and CLAP score are negatively correlated: if the balancing favors the text condition the CLAP score is higher, if it favors the audio condition the $\text{KNN}_\text{common}$ is higher. The double CFG thus works as expected. 

\section{Conclusion}

In this paper we introduced style conditioning for language model based music generative models: given a few seconds of a musical excerpt, one can generate music in the same style using our proposed audio encoder with an information bottleneck. We introduced new metrics to assess the equilibrium between generating music that maintains a similar style to the condition while also being different. We validated those with human studies. Finally, we can also mix this style conditioning with inconsistent textual description and balance them thanks to a new double classifier free guidance method. This method could be applied in other generative models with multiple conditions. 

\noindent\textbf{Ethical statement:} Improving music generation brings ethical challenges. Through carefully chosen bottlenecks in our style extractor (RVQ, downsampling) we aim for the right balance between increasing the model controllability and possible creative use while ensuring the model does not copy existing works, and provided new metrics to measure this. Finally, we only used music we licensed.

\clearpage
% For bibtex users:
\bibliography{ISMIRtemplate}

\end{document}